\documentclass[12pt,preprint]{aastex}
\usepackage{graphicx}
\usepackage[stable]{footmisc}
\usepackage{subfigure}
\shorttitle{Multi-band Receiver}
\shortauthors{The MBR Team}
\begin{document}
\title{RRI-GBT Multi-Band Receiver: Motivation, Design \& Development}
\author{
Yogesh Maan\altaffilmark{1,2}, 
Avinash A. Deshpande\altaffilmark{1}, 
Vinutha Chandrashekar\altaffilmark{1}, 
Jayanth Chennamangalam\altaffilmark{1,3},
K. B. Raghavendra Rao\altaffilmark{1},
R. Somashekar\altaffilmark{1},
Gary Anderson\altaffilmark{4},
M. S. Ezhilarasi\altaffilmark{1},
S. Sujatha\altaffilmark{1},
S. Kasturi\altaffilmark{1},
P. Sandhya\altaffilmark{1},
Jonah Bauserman\altaffilmark{4},
R. Duraichelvan\altaffilmark{1},
Shahram Amiri\altaffilmark{1,5},
H. A. Aswathappa\altaffilmark{1},
Indrajit V. Barve\altaffilmark{6},
G. Sarabagopalan\altaffilmark{1},
H. M. Ananda\altaffilmark{1},
Carla Beaudet\altaffilmark{4},
Marty Bloss\altaffilmark{4},
Deepa B. Dhamnekar\altaffilmark{1},
Dennis Egan\altaffilmark{4},
John Ford\altaffilmark{4},
S. Krishnamurthy\altaffilmark{1},
Nikhil Mehta\altaffilmark{1,7},
Anthony H. Minter\altaffilmark{4},
H. N. Nagaraja\altaffilmark{1},
M. Narayanaswamy\altaffilmark{1},
Karen O'Neil\altaffilmark{4},
Wasim Raja\altaffilmark{1},
Harshad Sahasrabudhe\altaffilmark{1,8},
Amy Shelton\altaffilmark{4},
K. S. Srivani\altaffilmark{1},
H. V. Venugopal\altaffilmark{1},
Salna T. Viswanathan\altaffilmark{1}
}
\altaffiltext{1}{Raman Research Institute, Bangalore, India}
\altaffiltext{2}{Joint Astronomy Programme, Indian Institute of Science, Bangalore, India}
\altaffiltext{3}{Department of Physics, West Virginia University, PO Box 6315, Morgantown, WV, USA}
\altaffiltext{4}{National Radio Astronomy Observatory, PO Box 2, Green Bank, WV, USA}
\altaffiltext{5}{School of Physics \& Astronomy, University of Manchester, Oxford Road, Manchester, UK}
\altaffiltext{6}{Indian Institute of Astrophysics, Bangalore, India}
\altaffiltext{7}{Department of Electrical Engineering, The Pennsylvania State University, Pennsylvania, USA}
\altaffiltext{8}{Department of Physics, Purdue University, West Lafayette, IN, USA}
\begin{abstract}
We report the design and development of a self-contained multi-band
receiver (MBR) system, intended for use with a single large aperture
to facilitate sensitive \& high time-resolution observations 
simultaneously in 10 discrete frequency bands sampling a wide 
spectral span (100-1500 MHz) in a nearly log-periodic fashion.
The development of this system was primarily motivated by need
for tomographic studies of pulsar polar emission regions. Although
the system design is optimized for the primary goal, it is also
suited for several other interesting astronomical investigations. 
The system consists of a dual-polarization multi-band feed (with
discrete responses corresponding to the 10 bands pre-selected as
relatively RFI-free), a common wide-band RF front-end, and 
independent back-end receiver chains for the 10 individual sub-bands. 
The raw voltage time-sequences corresponding to 16 MHz bandwidth each
for the two linear polarization channels and the 10 bands, are recorded 
at the Nyquist rate simultaneously. We present the preliminary results 
from the tests and pulsar observations carried out with the Green Bank 
Telescope using this receiver. The system performance implied by these 
results, and possible improvements are also briefly discussed.
\end{abstract}
\keywords{Instrumentation: polarimeters -- Instrumentation: miscellaneous 
-- Methods: observational -- pulsars: general -- 
pulsars: individual (B0329+54) -- pulsars: individual (B0809+74)}
\section{Introduction}
Even after several decades of pulsar studies, we remain far from
being able to relate the puzzling rich details exhibited 
by pulsars to the physical processes responsible for the observed 
radio emission. While the average pulse profile with polarization 
information reveals the viewing geometry \citep{RC69}
and gives some clues about the possible emission mechanisms
\citep[for example, see][and other papers in the same series]{Rankin83a},
the fluctuations from pulse to pulse seem to be more crucial and 
promising in probing the underlying physical processes
\citep[see, for example][]{DC68,Backer73,DR99}. Over the past decade,
studies of such fluctuations 
in pulsar signals suggest that plasma processes responsible for 
pulsar emission may be organized into a system of columns seeded by 
a carousel of ``spark'' discharges in the acceleration region 
and undergoing steady circulation around the magnetic axis. 
\par
\begin{figure*}
  \begin{center}
    \includegraphics[width=\textwidth]{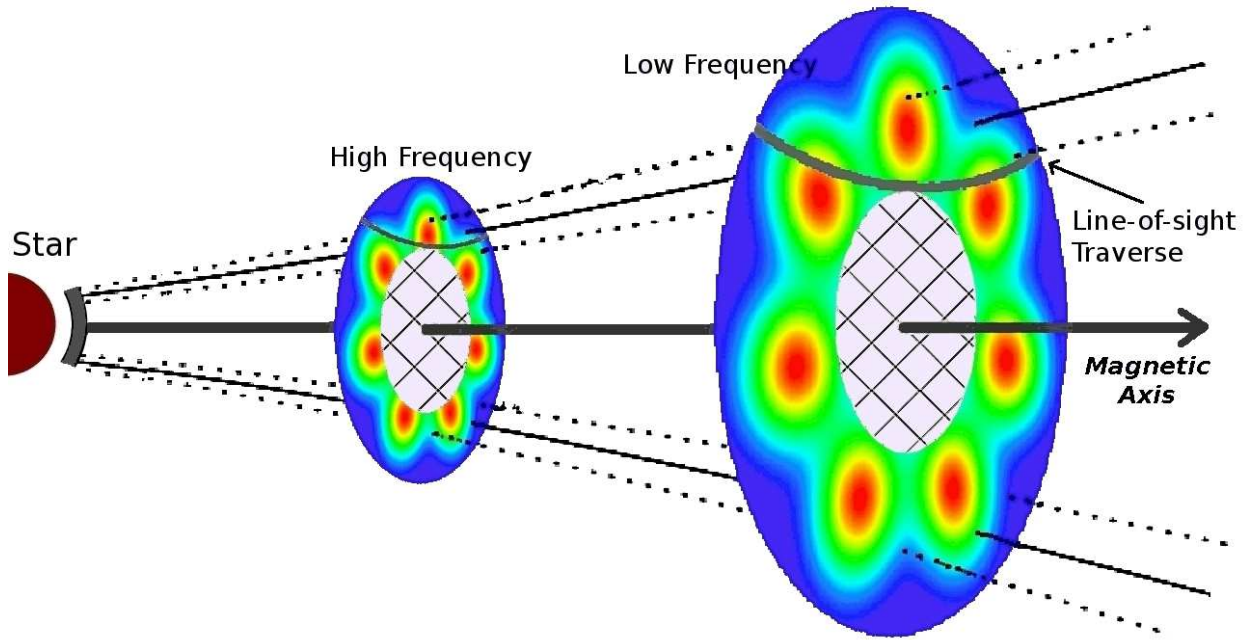}
    \caption{ A cartoon to illustrate tomography of polar emission 
region, wherein two of the many possible slices of the emission 
cone are shown consistent with the radius-to-frequency mapping. 
The gray arcs indicate the line-of-sight traverse.
Given the single pulse time sequences, 2-D pictures similar 
to those shown in the example slices, can in principle be mapped
using a cartographic transform \citep[for details, see][]{DR01}.
The size of the inner inaccessible region (cross-hatched) is determined by 
how close our sight-line gets to the magnetic axis. 
}
    \label{fig_cartoon0}
  \end{center}
 \end{figure*}

Estimation 
of the carousel circulation time, and hence ``cartographic'' mapping of the 
observed emission to a system of emission columns has been possible 
so far only for a limited number of pulsars, and the important factors 
determining such configurations and the pattern circulation are yet to 
be fully understood. Further, any possible connection between the
profile/polarization mode changes, as well as nulling, observed in 
several pulsars 
and the circulating patterns also needs careful investigations. 
To enable these investigations, ideally, we need full polarimetric 
data on how the 
emission columns and their configuration evolves with emission height 
and time. This demands simultaneous single-pulse observations across 
a wide frequency range implied by the so called radius-to-frequency 
mapping; with the high frequency emission originating close to the 
star surface, and regions of lower frequency emission progressively 
farther away. 
However, most of the available observations are made at a time over 
a narrow spectral range, thus sampling the details only in 
a slice of the emission cone. The limitation largely comes from
considerations of radio frequency interference (RFI), even if 
wide-band systems were to be available.
The need for a setup enabling 
multi-frequency observations, essential to sample the 
polar magnetosphere at various heights simultaneously, can not
be overstated. With such observations, polar emission mapping at a number 
of frequencies would not only allow ``tomography'' of the polar
emission region (see an illustration in Figure~\ref{fig_cartoon0}), 
but also might provide much needed clues about generation of sub-beam 
patterns, associated polarization and their evolution across the magnetosphere. 
\par
While in principle, simultaneous multi-frequency observations 
can be carried out using many telescopes simultaneously, 
practical difficulties in this approach are numerous. Apart from 
inherent complexities in coordinating/scheduling several telescopes
at different geographical locations, 
their available setups might have undesirable differences in, say,
hour angle coverage, aperture shape \& filling factor, polarization 
response, ionospheric contributions, synchronization (time-stamp) 
accuracy, calibration schemes, sensitivity, RFI-free bands and data formats. 
A superior alternative which provides the desired 
simultaneous multi-frequency observations,
avoiding all the above complexities,
is to have a ``self-contained'' multi-band system 
for use with a single large aperture. Such a system would
include a suitable feed, broadband front-end, parallel analog \& 
digital receiver pipelines,
along with appropriate monitoring, synchronization and data
recording systems. It is beneficial to sample
the signal voltages in dual polarization at each of the multiple
bands, pre-selected to avoid RFI as much as possible.
\par
We present here the design and development  of 
such a self-contained multi-band receiver system (MBR) for use
with the Green Bank Telescope (GBT), and describe preliminary
results from observations using this setup.
\par
Section 2 below gives a detailed hardware 
description of the system, section 3 describes the control and
monitor module. In section 4, we present the preliminary results
from the tests \& observations at GBT.
We discuss the system performance and possible future developments
in the concluding section.
\section{MBR System: Hardware Details}
\begin{figure*}
  \begin{center}
    \includegraphics[height=\textheight]{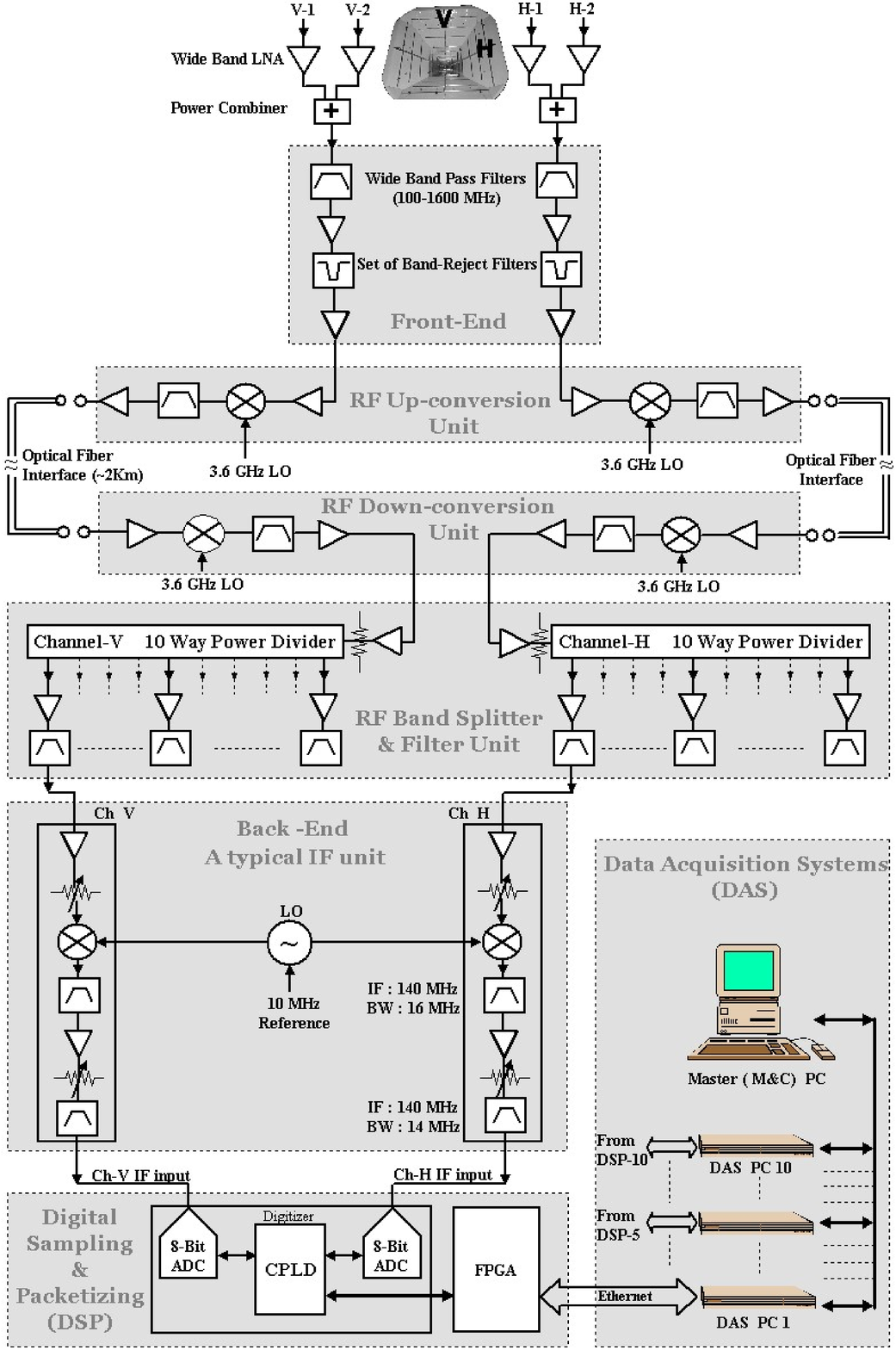}
    \caption{ The overall block diagram of the MBR.}
    \label{fig_block_diagram}
  \end{center}
 \end{figure*}

The overall block diagram of the MBR system, designed and 
developed at the Raman Research Institute (RRI), is shown in 
Figure~\ref{fig_block_diagram}. The signal path begins with a
dual linear polarization multi-band feed sensitive to 10 discrete bands
within a spectral span of 100-1600 MHz, followed by appropriate
pre-amplification, filtering, up-conversion, transmission over the
optical fiber, down-conversion, splitting in to bands of interest,
superheterodyning to intermediate frequency (IF), digitization 
and recording. Below we describe various parts of the MBR system
along this signal path.
\subsection{Multi-Band Feed}
Broadband feeds spanning more than a decade in spectral range 
are not commonly used in radio astronomy. Not only because their
designs are challenging, their utility in practice is severely
compromised due to ever increasing RFI from communication signals
occupying a large fraction of the radio band at frequencies below 2 GHz.
Low level RFI can be detected and excised during post-processing
of the recorded data. 
However, the presence of strong RFI in certain bands not only
makes the specific bands unusable, but it may also contaminate the
otherwise cleaner regions due to possible non-linear system response at
high power levels.
It is desirable therefore to reject RFI as early in the signal path 
as possible. Noting the challenges in designing feeds with uniform 
performance over a wide-band and the mentioned RFI issue, we have 
considered a multi-band design with good response only over selected
bands to provide preliminary immunity against RFI. 
As will be described later, the bands with severe 
RFI are filtered out explicitly soon after the signals from the feed are 
pre-amplified adequately (so as to minimize the effect of filter insertion 
losses in the desired band).
\par
\begin{figure*}
  \begin{center}
    \includegraphics[width=\textwidth]{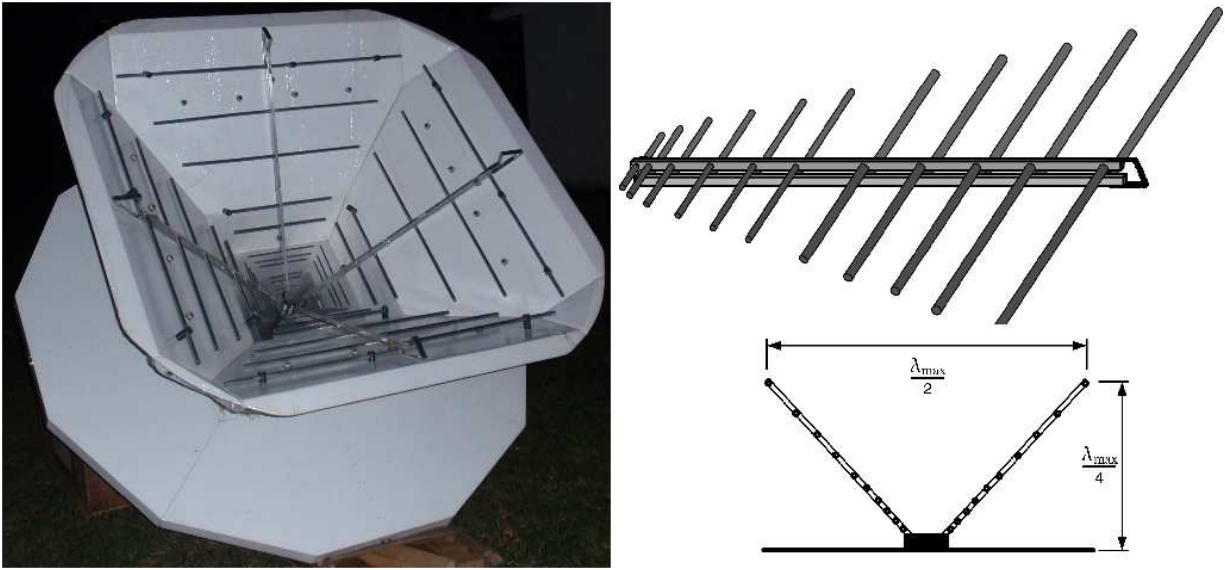}
    \caption{ Multi-Band Feed Schematics; Left: A photograph of the 
multi-band feed installed at the primary focus of the GBT. Right: 
Upper part shows the position arrangement of dipoles 
for one half of the single polarization feed, lower part shows the 
lateral view of the single polarization feed-element.}
    \label{fig_feed}
  \end{center}
 \end{figure*}

For a broadband feed antenna at the prime focus of a parabolic dish, 
it is important to ensure that the phase center of the feed remains
independent of frequency. Also desirable is a good match between the
E \& H plane responses, while using linear feeds or their combinations
to illuminate circular apertures uniformly. In view of these, we find
the basic conical arrangement of active elements as in the ``Eleven Feed'' 
\citep{Olsson06} to be most suited for our purpose. In these feeds,
log periodic dipole-arrays are arranged in pairs at $45^{\circ}$ 
about the symmetry axis for each one of the two linear polarizations
(constraining the apex angle of each of the dipole-arrays to $35^{\circ}$).
In the Eleven feed the phase center does not move significantly with 
frequency, and the spectral response is rather uniform across the
design band. Our design differs significantly from that of the Eleven feed 
specifically in the latter of these two aspects. As shown in 
Figure~\ref{fig_feed},
we use a plane reflector (ground plane, 1.7 m across) perpendicular to the 
axis of the conical structure. 
For each one of the dipoles in the four dipole-arrays, when considered 
together with its corresponding image, the combined phase center would
be at the symmetry point in the plane of the reflector. Extending this
picture to two dipoles of a given frequency, corresponding to a pair of 
dipole arrays for one polarization, it is easy to see that the phase
center of the combination will be at the apex of the cone in the
reflector plane. More importantly, this would be independent of
frequency, and in principle also independent of the opening angle of 
the cone. 
Each of the dipole arrays following an approximately
log-periodic pattern, consists of a set of resonant dipoles corresponding
to the bands of interest, chosen as the relatively RFI-free parts
of the wide-band.
\par
A number of multi-band feed designs, differing in dipole arrays, their
mounting and feed-point connections, were experimented with, keeping
the basic conical arrangement and the ground reflector as mentioned above. 
Details of a couple of the early designs, and their performances can 
be found in \citet{Maan09}.
The final design of the dipole array (Figure~\ref{fig_feed}) consists 
of 10 simple half-wave dipoles arranged in a quasi-log-periodic fashion,
along with a stub terminating the feed-line past the longest wavelength
dipole and an extra shorter dipole past the highest frequency resonant
dipole. 
The dipoles are realized using aluminium tubes and rods at frequencies
below and above 300 MHz respectively, with diameters progressively
decreasing with wavelength as far as possible. The central feed-line
(of length 73.5 cm) 
consists of a pair of 10 mm aluminium square rods and the stub at its
end made of a 10 mm wide, 2 mm thick aluminium strip. The length of the
stub was optimized based on a combined criteria of favorable and poor
response of the feed-line (modelled as a transmission line with one 
end shorted) at the desired and the RFI bands respectively. The value 
obtained from a computer program to search for the optimum length
matched well with that assessed from trial \& error. The support 
structure for the dipole arrays (conical section behind the dipoles 
in Figure~\ref{fig_feed}) is made of fiber glass material, keeping it
light-weight but sturdy. The dipoles are secured on this structure
with the help of Nylon bolts and the face of the cone is covered
with Tecron cloth.
\par
The presence of dielectric behind the dipoles does, in principle,
affect their performance, and increasingly so at the higher frequencies
($>500$ MHz). Hence, the
conical profile was maintained only till the 430 MHz dipole.
In the 
lower part of the structure, the close proximity to (higher frequency) 
dipoles is avoided by the profile deviating from the conical to a 
boxy profile at the base.
The resonant dipole at the lowest frequency would have been too long
to accommodate within the size constraints at the prime focus of the GBT.
Hence, a shorter length dipole is used, with its location on the feed-line
appropriate to its actual length, thus reducing the overall size of the feed.
The feed-line is designed with a characteristic impedance of 50 $\Omega$, 
and a co-axial cable is directly connected to the feed-points of each of 
the four dipole arrays. These cables are kept short in length, just enough
to reach the pre-amplifiers located immediately below the reflector. 
\par
\begin{figure*}
  \begin{center}
    \includegraphics[width=0.5\textwidth]{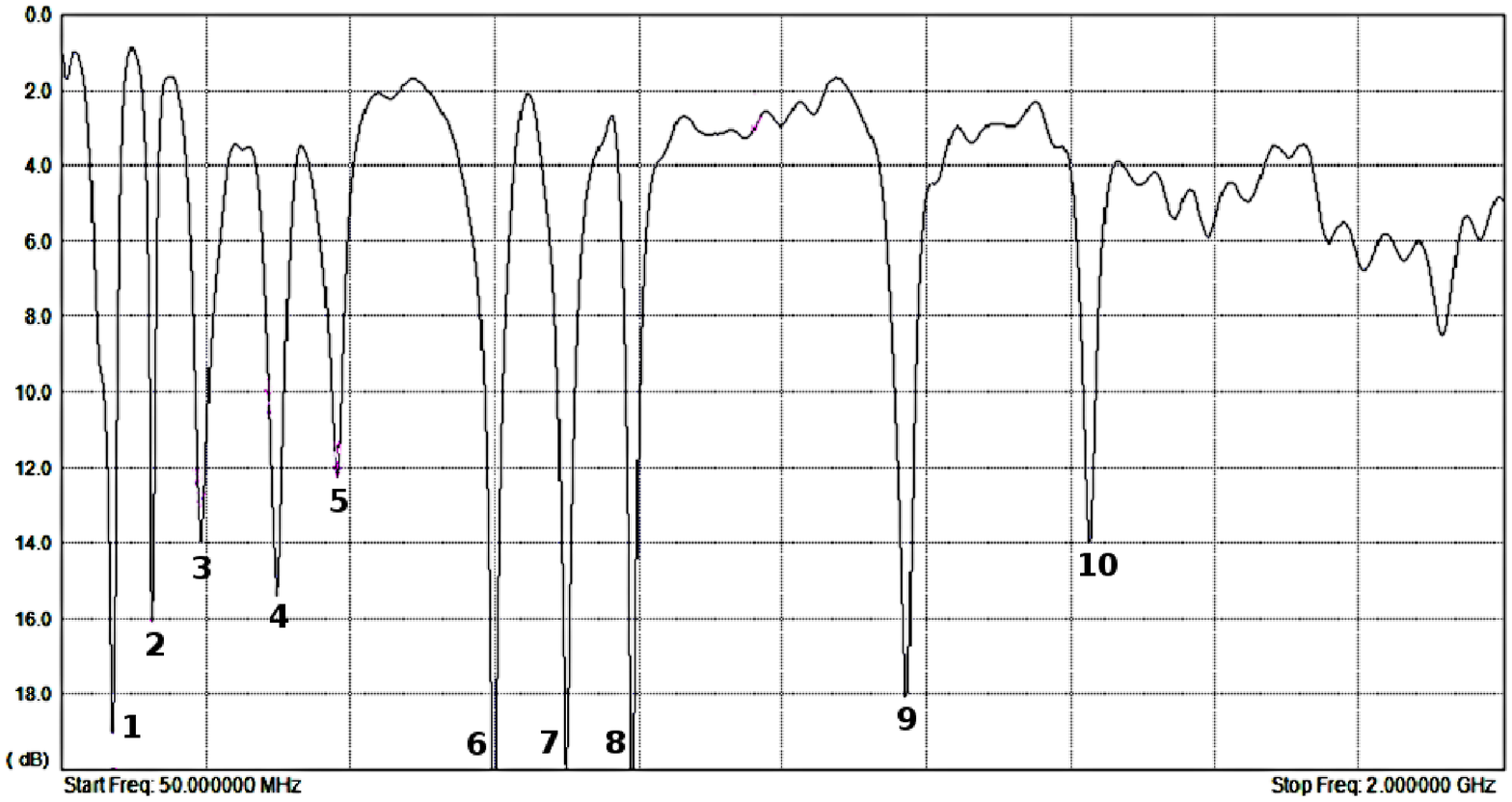}
    \caption{Multi-Band Feed response in terms of the return-loss profile,
measured using one of the polarizations, is shown in the frequency range
50 MHz to 2.0 GHz. The vertical scale corresponds to a range of 0
to 20 dB. A return-loss of more than 10 dB in about $\geq 20$ MHz wide bands
centered at the 10 desired frequencies can be clearly seen. In most of the
\textit{unwanted} part of the spectrum, the return-loss is constrained to
less than 4 dB.}
    \label{fig_feed_response}
  \end{center}
 \end{figure*}

The in-phase combination of the two dipole arrays at angles 
$\pm45^{\circ}$ about the feed axis and corresponding to a
given linear polarization, provides a good match between the
E \& H plane beams. For a beam with full width at half power of
about $60^{\circ}$, and the half angle of $42^{\circ}$ for the 
GBT dish as seen from its prime focus, the edge taper would 
be at least -7 dB, before including the inverse square law dependence
on the distance from the feed. This would amount to relatively larger spill 
over than usual, but would illuminate a larger effective collecting 
area. In comparison, the edge taper inferred from the beam-width, as we
will see later, is somewhat larger (in the range -15 dB to -20 dB). 
Figure~\ref{fig_feed_response}
shows a typical return-loss profile across the
spectrum for one of the dipole array pairs (i.e. corresponding to one 
of the linear polarizations).
The feed cross-polarization is measured, both during the feed tests
and from observations of bright unpolarized sources, and is found to
be at the level of $-20\,dB$. 
\subsection{Wide-band RF Front-end}
As noted above, each of two dipole-array outputs corresponding to a
linear polarization are first amplified separately and then combined 
in phase. 
As a wide-band ($\le 2$ GHz) low noise amplifier (LNA), we have
used a commercially available module of Mini Circuits ZX60-33LN, with
a typical gain of around $20\,dB$ and noise-figure less 
than $1.3\,dB$ over the band of interest.
The pre-amplified signals, after the respective pairs are combined, 
provide dual orthogonal polarization outputs that are passed through
a band limiting filter (100-1600 MHz), realized using appropriate
combination of low pass and high pass filters. 
After further nominal amplification, the bands associated with known
strong RFI signals (e.g. FM radio signals, TV transmission
\& mobile communication and those particular to the geographical 
location of the system; presently GBT) within our primary band are 
rejected using a cascade of home-made band-reject filters. At this
point, since the primary contaminants are removed, larger amplification 
is possible and needed to carry the signals from feed location to
the receiver room. The modules in the receiver room prepare the signal
for subsequent transmission to the equipment room via 
existing optical fiber system. These modules up-convert the
RF band of 100-1600 MHz to 2000-3500 MHz, using 
a local oscillator (LO) of 3.6 GHz. Noting that the band is flipped 
in this translation, appropriate length RF cable is introduced
in the path (in order to use its frequency dependent attenuation 
characteristics) to at least partly compensate for the effect of 
gain-slopes introduced in the earlier stages.
In the equipment 
room, the received band is down-converted back to the original RF band 
of 100-1600 MHz, using an LO again of 3.6 GHz.
\subsection {Sub-band Receiver Chains}
The wide-band RF signal received in the equipment room is split into 
10 fixed sub-bands\footnote{Although the system was originally conceived 
to provide operation with 8 sub-bands, 2 additional back-end chains built
as spares were also  made part of the system providing 2 extra sub-bands.}
after adequate amplification in anticipation of the power-divider losses.
The bandwidths of the filters separating the different bands varies
typically between 20 to 120 MHz, decided based on the RFI 
situation\footnote{An RFI survey at the observing site 
was performed primarily around the desired sub-bands, and the nearest 
available relatively RFI-free regions were selected.}. 
The details of the central radio frequency values and the corresponding 
bandwidths of the 10 RF sub-bands are given in Table~\ref{table_filters}. 
The chosen bands provide nearly log-periodic sampling within the wide-band.
\subsubsection{Analog Pipeline}
The 10 signals, thus obtained, are processed through very similar
high gain chains, each consisting of an amplifier-attenuator combination 
for RF, a mixer followed by a filter, an amplifier-attenuator combination 
\& another band-defining filter for IF, in that sequence. 
The LO signal, required for the RF to IF conversion, is generated using 
a discrete PLL (phase-locked-loop) wherein a VCO (voltage controlled 
oscillator) is phase-locked to a 10 MHz signal from a rubidium standard,
which in turn is disciplined with a 1PPS (pulse per second) derived from
the Global Positioning System (GPS). 
The LO frequency can be selected in steps of 1 MHz, within a range
large enough to tune in to any region of the respective RF sub-bands.
Care is taken in choice of the IF frequency to ensure that the IF
as well as any image band falls outside the respective RF sub-band.
Thus, the IF center frequency for most of the chains is 140 MHz,
while it is 70 MHz for the two lowest RF bands (below 200 MHz).
The IF-filters are commercial surface acoustic wave (SAW)-filters 
of 16 MHz bandwidth, providing a sharp roll-off characteristic and 
good rejection outside the IF band. The second of the IF-filters is
necessary to suppress out-of-band noise-floor contributed by the
electronics following the first of the IF-filters.
The independent attenuators in the RF and IF sections, which are computer
controlled and provide attenuation within 0 to 31 dB in steps of 0.5 dB,
are used to adjust the signal level to ensure operation in the linear
regime of the chain. The net amplification in the signal path is 
arranged to be typically in excess of 100 dB so as to provide  a 
suitable level of IF input to the analog-to-digital converter (ADC).
\subsubsection{Digital Section}
\begin{figure*}
  \begin{center}
    \includegraphics[width=\textwidth]{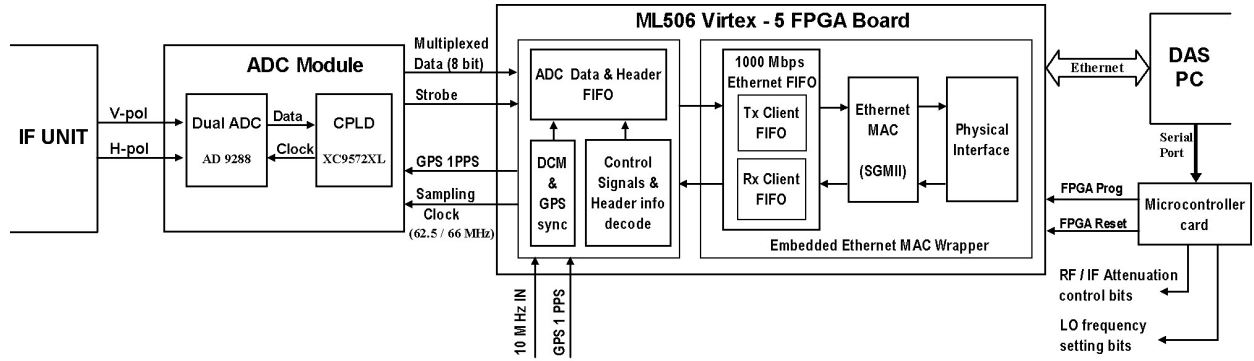}
    \caption{Block diagram of the ``Digital Section'' in a typical sub-band receiver chain.}
    \label{fig_fpga}
  \end{center}
 \end{figure*}

The IF signals for the two polarizations corresponding to a given 
sub-band, with 16 MHz bandwidth \& centered at 140/70 MHz, are
digitized directly using harmonic sampling. 
A low-power, 8-bit 
dual-channel ADC (AD9288) capable of operating 
at 100 mega samples per second, and with an analog bandwidth of 475 MHz 
is used for this purpose (see Figure~\ref{fig_fpga}), providing high 
dynamic range\footnote{Tests with a CW input at frequency of 16 MHz 
to characterize the ADC indicated SNR of 32 dB, implying the Effective 
Number Of Bits (ENOB) to be 5.5 
for each of the two channels.}.
The sampling clock frequency in our case is 33 (or  31.25) MHz for 
IF bands at 140 (or 70) MHz.
The ADC is controlled by a CPLD (Complex Programmable Logic Device; 
XC9572XL), configured using VHSIC hardware description language (VHDL),
where an input clock at twice the sampling frequency is de-multiplexed 
and fed to the two channels of the ADC. At the start of each acquisition,
the clocks are synchronized with respect to 1 PPS signal 
generated from GPS-disciplined rubidium standard. The dual channel 8-bit
samples along with strobe are time multiplexed and passed to the following
module.
\par
The digital back-end is realized using a Xilinx-ML506 evaluation board
consisting of a Virtex-5 FPGA (field 
programmable gate array), and a rack-mountable PC. The 8-bit 
parallel input stream at the rate of about 66 mega-samples per second,
representing multiplexed raw voltage samples from the two orthogonal
polarization channels, is received and after appropriate packetization
the data are transmitted to the PC on a Gigabit Ethernet (GbE) channel.
The FPGA firmware for this purpose has been 
developed in VHDL. The GbE interface logic is realized on the FPGA 
using the intellectual property (IP) core ``Embedded Tri-Mode Ethernet 
MAC Wrapper v1.3'' provided by Xilinx. The basic block diagram of the 
digital back-end receiver is shown in Figure~\ref{fig_fpga}.
The on-board FPGA is programmed through a serial port on the PC via 
a micro-controller. Initialization is performed through a master reset
bit generated by control software on the PC. The functionality in
the FPGA includes generation of both the sampling clock (66/62.5 MHz, 
used by ADC) and a 100 MHz clock used for reading the ADC data buffered
through a FIFO (First In First Out). These programmable clocks are 
generated from an input reference clock at 10 MHz (from the already
mentioned rubidium standard) using the digital clock manager (DCM)
blocks in the FPGA. A GPS-disciplined rubidium-based 1PPS 
input to the FPGA is used for both, synchronization at the start
of the acquisition as well as for its time-stamping. Several pieces
of information relevant to receiver settings and acquisition
details (sent from the control software on the PC) are decoded, 
using a separate firmware, before their inclusion in the packet header. 
\par
\begin{figure}
  \begin{center}
    \includegraphics[width=0.5\textwidth]{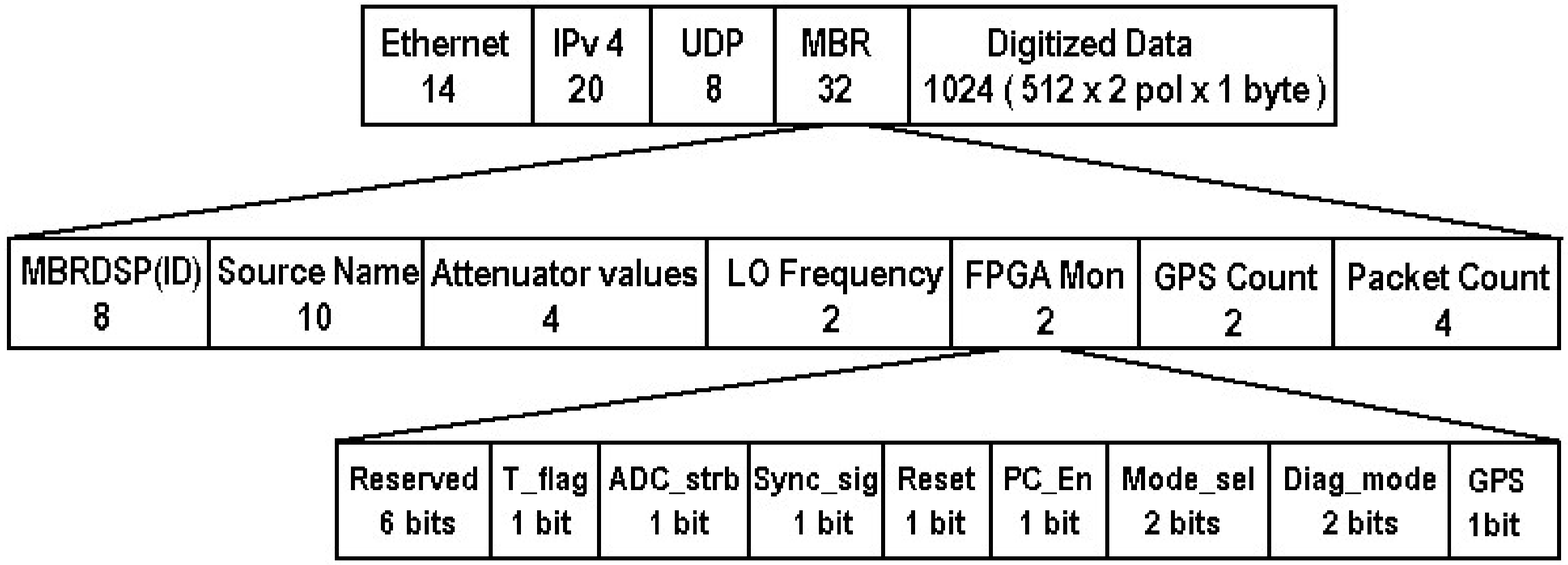}
    \caption{ Ethernet Data Packet Structure (The numbers shown in the 
first two rows are in terms of bytes, as opposed to bits in the third row).}
    \label{fig_frame}
  \end{center}
 \end{figure}

The FPGA logic is designed to transmit the data and header information
in packet form, using the user datagram protocol (UDP). 
Each data packet consists of 42 bytes of 
protocol header (Ethernet, IP and UDP headers), 32 bytes of MBR 
header (which contains the identity of the receiver, status information 
like mode of operation, GPS counter, and packet counter), and 1024 
bytes of digitized data (dual-polarization, interleaved), as shown in 
Figure~\ref{fig_frame}. 
Any change made in the system settings is reflected instantaneously in 
the headers associated with the acquired data (i.e. in the data-packet 
immediately following the change).
\section{MBR System: Software Details}
The monitoring \& control of the MBR as well as data acquisition are 
handled by a specially developed software suite.
The 10 rack-mountable PCs, referred to as data acquisition
subsystems (DAS), interconnected through a GbE-switch and coordinated 
by a master monitoring-and-control (M\&C) PC, together form a part of
the digital back-end, and provide a platform on which the relevant
codes are executed.
\par
\begin{figure}
  \begin{center}
    \includegraphics[width=0.5\textwidth]{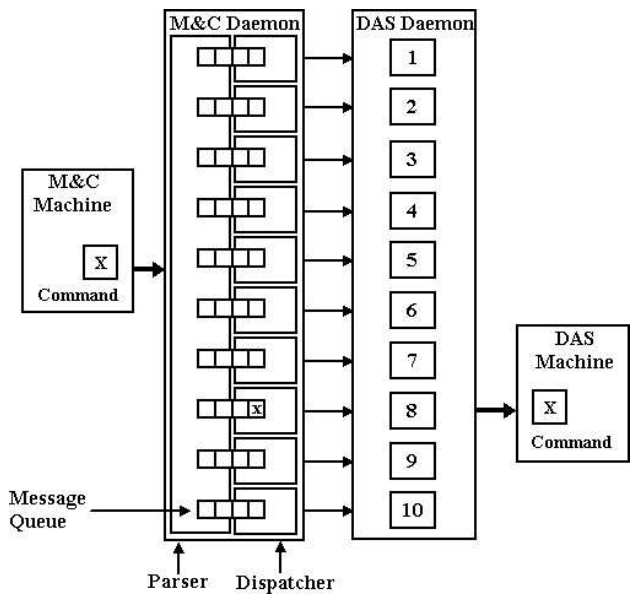}
    \caption{Software architecture of the MBR system. Here, `X' is a command
             intended for DAS 8. The parser unit of the M\&C daemon pushes
             this command into the message queue for the eighth dispatcher,
             which in turn sends the command over the network to the DAS
             daemon. The DAS daemon runs the binary corresponding to
             the command.}
    \label{fig_softarch}
  \end{center}
\end{figure}

The software that runs on the MBR system consists of three major 
components -- commands, the M\&C daemon and the DAS daemons, the latter
two running on the respective machines.
Commands are used to 
set and retrieve system settings, as well as
to perform specific tasks, such as triggering data acquisition.
The M\&C daemon acts as a 
command router -- it receives user commands and dispatches them 
to the relevant DAS machine, where the local DAS daemon executes them.
The M\&C software system architecture, depicting a sample 
command flow, is shown in Figure~\ref{fig_softarch}.
\par
The command-based control 
facilitates (a) setting of RF/IF attenuation and LO frequency relevant to
the analog section, and (b) GPS synchronization, specifying bit-length/sample,  
duration and mode of acquisition, etc. for the digital back-end, along with
initiation and termination of data acquisition.
When triggered, instances of 
the data-acquisition program run on the DAS machines, each recording
packeted data (received through a GbE link from the digital back-end) 
to the local disk at full rate, as well as sending `sniff-mode' data 
at a much reduced rate (typically, 1 out of every 1000 packets) 
to the M\&C machine for monitoring and recording for
subsequent processing.
\par
\begin{figure*}
  \begin{center}
  \subfigure[]{
   \includegraphics[angle=-90.0,width=0.47\textwidth]{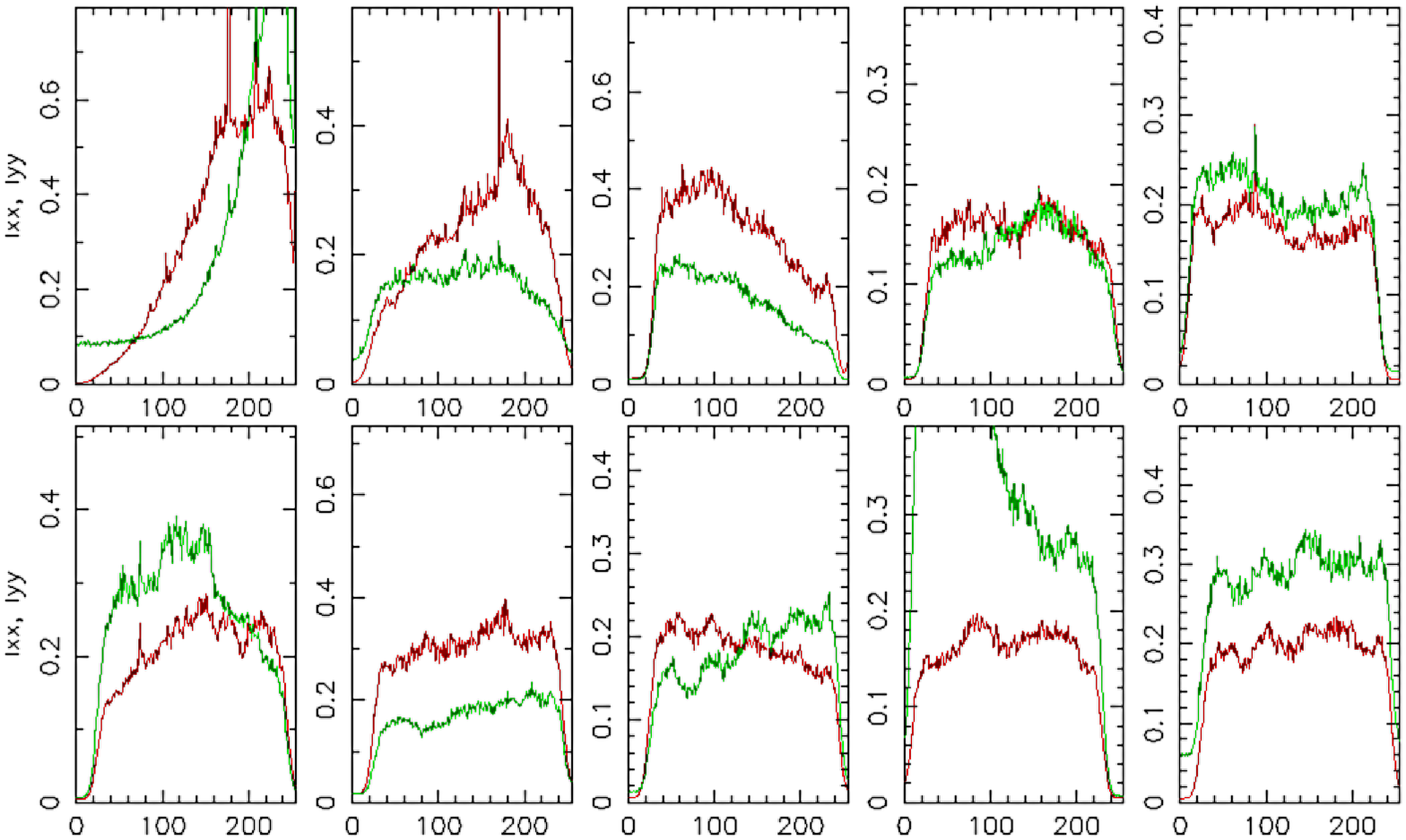}%
   \label{fig_monitor_spec}
   }
  \subfigure[]{
   \includegraphics[angle=-90.0,width=0.47\textwidth]{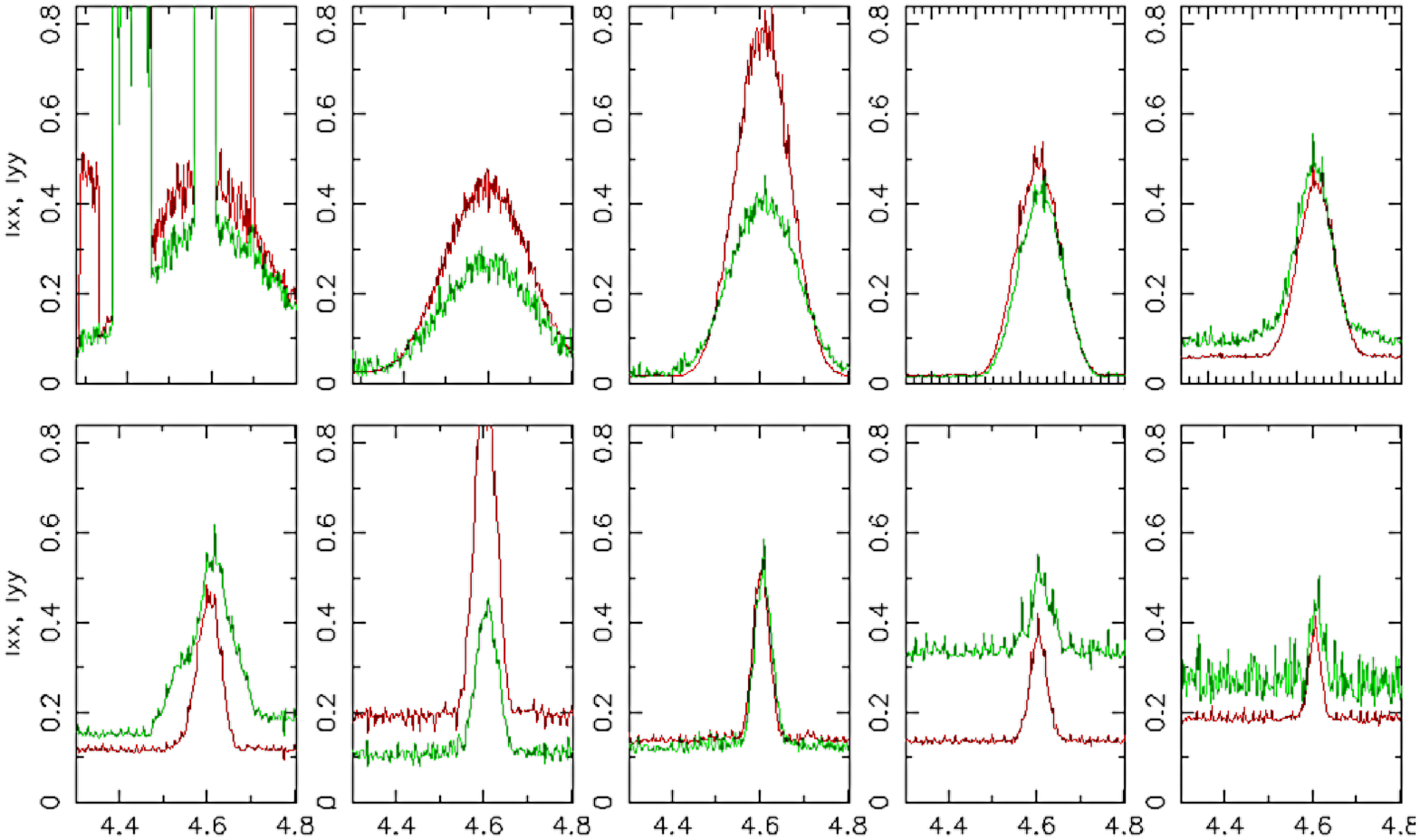}%
   \label{fig_monitor_time}
   }
  \caption{A snapshot of spectra as well as total power time-sequences
being monitored in quasi-real-time, simultaneously in the 10 frequency
bands, is shown here. Panel (b) shows total power scan-data of slewing 
across the bright source Cas-A, while panel (a) shows the spectra corresponding 
to the last averaged samples of the time-sequences shown in panel (b). 
The two curves in each of the
sub-panels correspond to the two orthogonal polarizations. Each one of the
spectra consists of 256 channels across 16 MHz of bandwidth, and the total 
extent of the horizontal axes shown in panel (b) corresponds to about half 
a minute.}
  \label{fig_monitor}
  \end{center}
\end{figure*}

Checking for the expected
pattern in the packet counter values, available in the headers of
packets received on the M\&C machine,
serves as a quick diagnostic of acquisition on the respective 
DAS machines. 
Processing of these sniffed data is performed by a separate software and 
quasi-real-time spectra (in full Stokes), as well as total power 
time-sequences, can be monitored
(as in Figure~\ref{fig_monitor}).
This software also facilitates estimation 
and automatic adjustment of the gains of individual receiver chains 
to equalize the output power levels, a tool often used before the 
start of acquisition. Spectral scans can also be performed, by stepping
through LO settings, to assess system gain and RFI occupancy across the
accessible spectral span, for all chains simultaneously.
\section{Observations and Preliminary Results}
After successful intensive laboratory tests, performance of the MBR
system was assessed through preliminary observations, 
wherein the broad beam of the multi-band feed was slewed 
across bright astronomical sources. These tests were conducted
at the RRI field station -- Gauribidanur observatory -- which provided
a relatively RFI free environment. 
To test the system including its use with a shaped reflector,
the feed was installed at the prime focus of one of the GMRT\footnote{Giant 
Meter-wave Radio Telescope} dishes 
\citep[see][for details]{Maan09}. 
Through the more sensitive 
measurements made possible by the large collecting area of the dish,
the performance of the feed and the system as a whole was characterized
in terms of aperture \& beam efficiencies, system temperature, etc.
These tests provided valuable feedback which prompted improvements in
the feed design and in the filtering strategy in the front-end.
\par
\begin{figure}
  \begin{center}
    \includegraphics[angle=-90.0,width=0.5\textwidth]{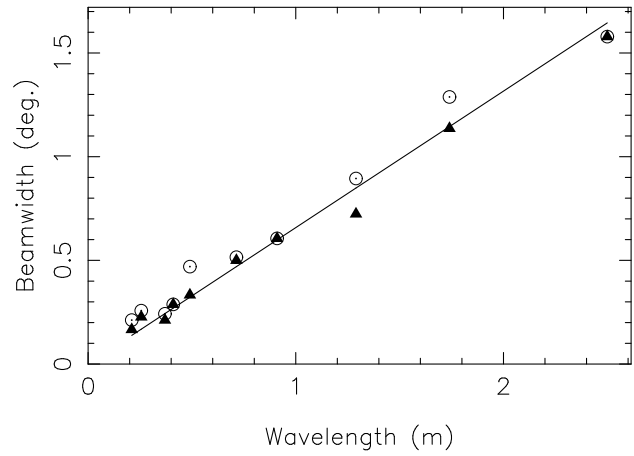}
    \caption{ Estimated beam-widths from Cas-A scans are examined
for linear dependence on wavelength (corresponding to the center frequencies 
listed in Table~\ref{table_filters}).
The circles \& triangles correspond to the estimates for the beams
associated with the two orthogonal linear
polarizations. The solid line indicates the best-fit linear trend
modelling the wavelength dependence of the observed beam-widths.}
  \label{fig_beamwidths}
  \end{center}
 \end{figure}

After these revisions,
the system was moved to Green Bank in May, 2009. A standard GBT 
receiver-box (used at its prime focus) was made available to house 
the MBR front-end as well as to attach the multi-band feed. After
the system was fully installed, test observations were conducted on 
a few bright continuum sources. These resulted in measurements of 
the beam patterns as well as helped in deciding the optimum location 
of the feed based on a focus scan. Attenuation/gain adjustments 
necessary to operate within linear regime of the optical fiber system 
at GBT were incorporated, and using spectral scans, regions of better 
sensitivity were identified for each of the bands.
Slew-scans across the bright radio source Cas-A (as shown in
Figure~\ref{fig_monitor_time}, obtained simultaneously in all the
10 bands) were used to estimate the
beam-widths (full widths at half power) corresponding to the two
polarizations. These estimates are shown as a function of wavelength 
in Figure~\ref{fig_beamwidths}.
As desired, the illumination of the GBT dish by the multi-band feed
appears uniform across frequency (reflected by the goodness of the 
linear fit). If interpreted assuming Gaussian illumination, the
illumination efficiency is about $75\%$. Given possible differences 
in the E \& H plane beams, and the different diameters of the GBT
dish in the orthogonal directions (i.e. 100 \& 110 meters), the 
implied edge-taper would be in the range -15 dB to -20 dB.
\par
\begin{figure}
  \begin{center}
    \includegraphics[angle=-90.0,width=0.5\textwidth]{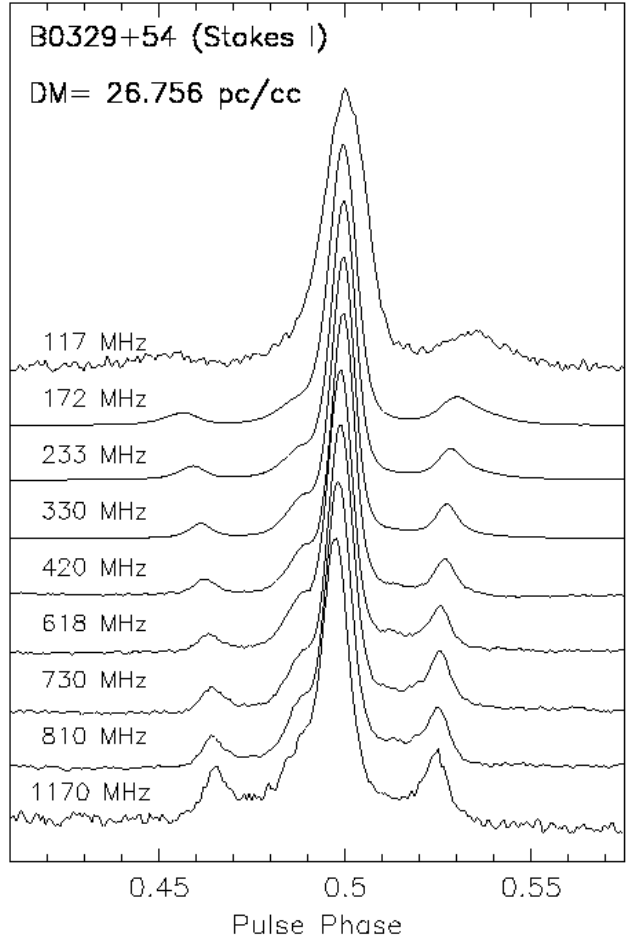}
    \caption{The average profiles of the bright pulsar B0329+54 obtained
from simultaneous observations in the 9 bands, after appropriate 
dedispersion, show spectral evolution of the pulse profile in total intensity.}
    \label{fig_avg_profs}
  \end{center}
 \end{figure}
\begin{figure*}
  \begin{center}
    \includegraphics[angle=-90.0,width=\textwidth]{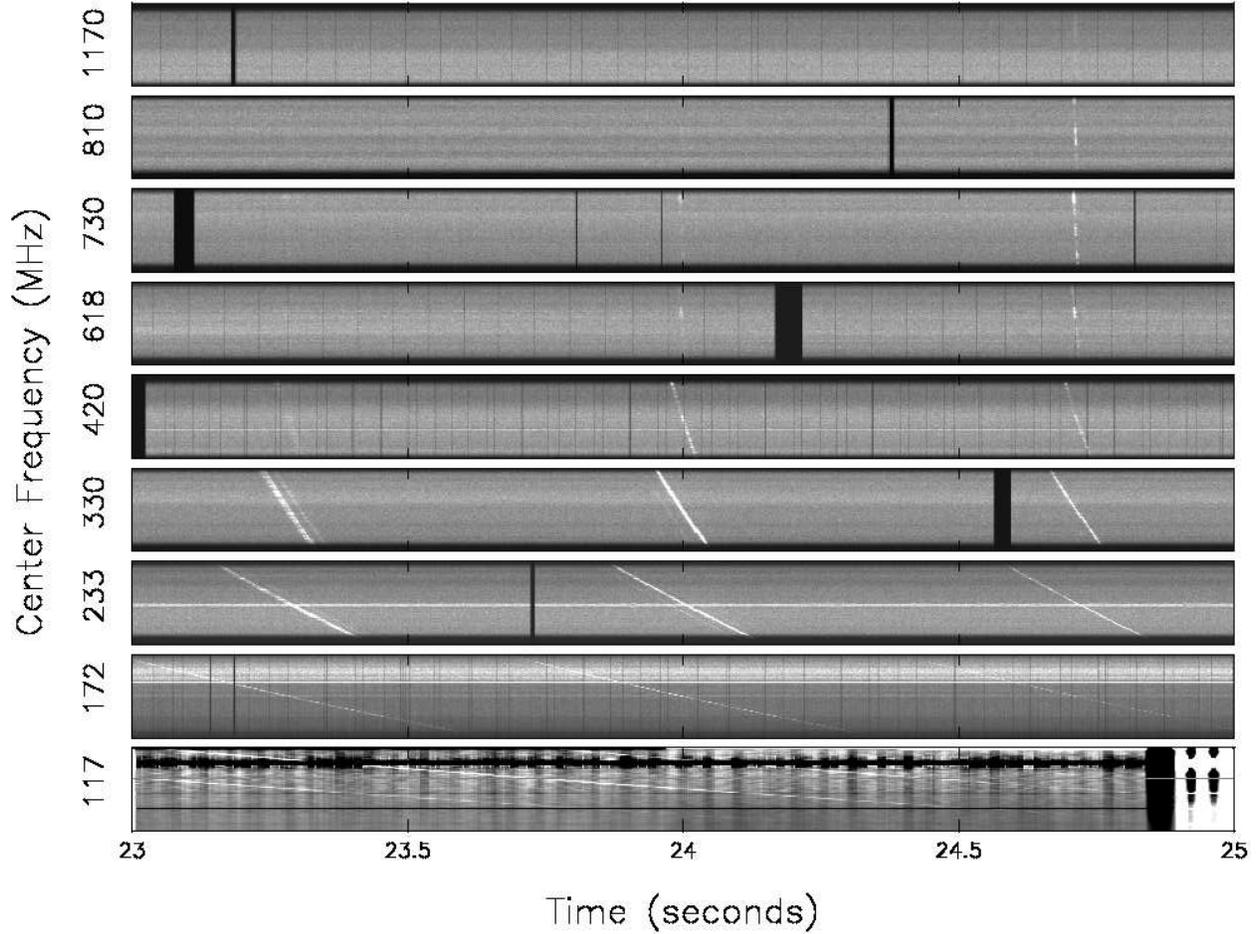}
    \caption{The dynamic spectra for each of the 9 bands
(stacked together), across a 2-seconds span, are shown here. 
The faint slanted \& curved streaks within the bands highlight the 
dispersed pulses, wherein the data across the
various bands have been shifted, removing the relative dispersion delays 
between the respective bands (referenced to respective center frequencies).
As expected, the slope of the streaks becomes shallow as we go to lower 
frequency bands with steep increase in the dispersion delays across the 
respective bands. The first band (centered at 117 MHz) is severely contaminated 
by strong wide-band (e.g. white patch between $\sim24.9-25$ seconds) as 
well as narrow-band RFI.}
    \label{fig_dyn_spec}
  \end{center}
 \end{figure*}

After the above tests, about 20 hours of observations were carried out 
on a number of bright pulsars, along with short duration pointings at 
Cas-A, Cygnus-A and Crab as flux calibrators. The results from these 
data (enabling investigation of single pulse fluctuations
over multiple bands simultaneously) and the related implications for the
underlying sub-beam pattern \& its evolution over emission height, etc.,
will be reported elsewhere. However, in this paper we conclude with
presentation of illustrative examples of the type of data obtained
with this MBR system.
Figure~\ref{fig_avg_profs} shows the phase-aligned average 
profiles of the pulsar B0329+54 observed in 9 bands of the 
MBR (the 10th band data were unavailable due to an unfortunate
computer crash). A sample of raw dynamic spectra in the corresponding bands, 
illustrating typical data quality (including RFI contamination)
across the respective bands, is shown in Figure~\ref{fig_dyn_spec}.
\par
From several pulsars observed to study drifting sub-pulses, we
use our B0809+74 observation to illustrate construction of a tomograph.
This pulsar exhibits remarkably stable pattern of drifting
sub-pulses across the pulse window \citep{VS70,Page73}.
Given the viewing geometry
and the emission pattern circulation period ($P_4$),
such fluctuations in the single pulse sequence can be transformed 
to reconstruct a rotating carousel of sub-beams \citep{DR99,DR01}.
Estimation of the circulation period might be possible directly, if
an associated spectral signature is detectable. 
Alternatively, the time-period between
the consecutive sub-beams (i.e. interval between the sub-pulse 
drift bands; $P_3$) combined with their total number ($N$) could also
be used to estimate $P_4$ = $N$.$P_3$.
\par
\begin{figure}
  \begin{center}
    \includegraphics[angle=90.0,width=0.5\textwidth]{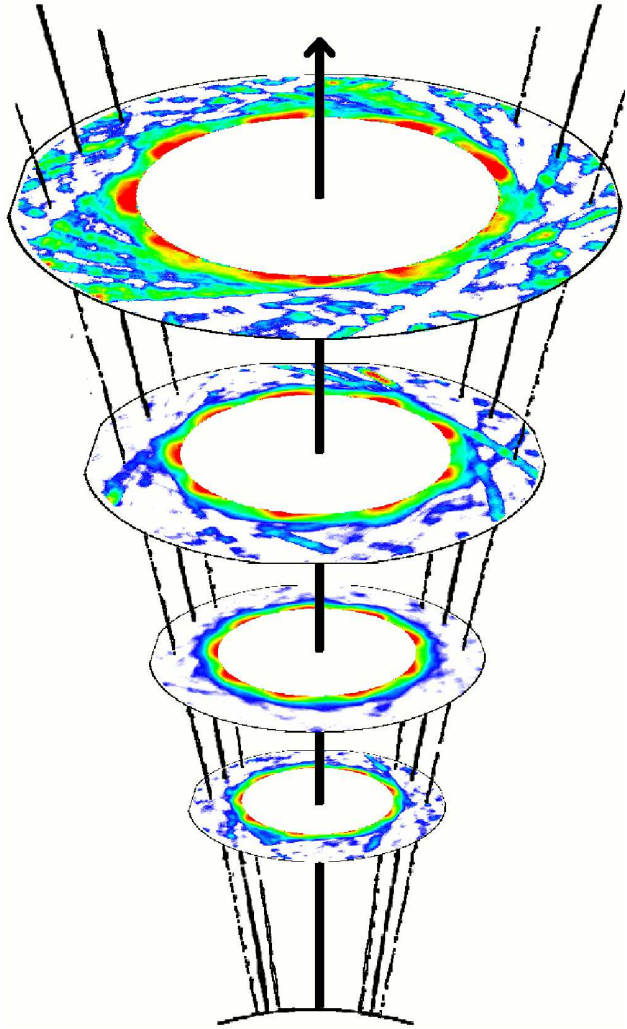}
    \caption{This preliminary tomographic view of the polar emission region 
of the pulsar B0809+74 is obtained by vertically stacking maps of 
the underlying emission patterns at the 4 lowest frequency bands 
(across 117-330 MHz), where each map is resulted from cartographically 
transforming the single pulse fluctuations observed at the corresponding 
radio frequency. The relative locations and sizes of these patterns, 
chosen conveniently, are consistent with the essential qualitative 
trends of radius-to-frequency mapping and geometry of the emission cone.
The overall similarity of the patterns across frequency, in terms of
the number of sub-beams (the high intensity regions corresponding to the 
red color) and their relative locations, is clearly evident. The partial 
sampling of the radial extent of the sub-beam is seen to improve
systematically towards lower frequencies, corresponding to the expected
broadening of the polar emission cone.}
    \label{fig_tomograph}
  \end{center}
 \end{figure}

In the present case, our fluctuation spectral analysis revealed 
a single high-Q feature in an otherwise featureless spectrum, 
providing an estimate of $P_3$ to be
$10.8529\pm0.0007$ spin periods\footnote{Although, no direct signature of
the circulation period is apparent in the fluctuation spectrum, 
the narrowness of the feature associated with $P_3$ argues 
strongly in favour of its integral relationship with $P_4$, and
also that the coherence in this drift sequence appears to
be largely unaffected by `nulling', if any.}.
Using the geometrical
parameters\footnote{magnetic inclination angle $\alpha=8.8^{\circ}$, and
sight-line impact angle $\beta=4.7^{\circ}$} from \citet{Rankin06},
and assuming a total of 9 sub-beams, 2048 periods long pulse sequences
are cartographically transformed to map the emission pattern 
in each of the sub-bands.
While further details of analysis on this pulsar and the assumed
carousel circulation period
would be discussed elsewhere, here we present a 
preliminary tomograph of the polar emission region of this pulsar
in Figure~\ref{fig_tomograph}, using data in the lowest 4 frequency
bands having adequate signal-to-noise ratio.
\section{Discussion and Conclusions}
The examples shown in previous section have illustrated 
the quality and the types of data obtained in simultaneous 
multi-frequency observations facilitated by the MBR.
The efficiency of illumination of the GBT reflector by our
feed appears quite uniform across the 10 bands, and our focus
scans taken on strong continuum sources during the commissioning 
tests indicate the feed phase-center to be independent of frequency.
Although the feed performance is as desired in these aspects, we 
find significant loss of sensitivity at the higher frequency bands
in general, and at the 1200 \& 1450 MHz bands in particular.
\par
Implicit in our feed design is the requirement that the two dipole 
arrays corresponding to a given linear polarization are phased 
correctly, particularly at the frequencies of interest.
This requires matching of spectral responses between the two 
arrays, and any mismatch in them would cause improper phasing
(since phase may vary rapidly about the respective resonant 
frequencies), and consequently significant de-phasing across 
the primary reflector that the feed illuminates. In practice, 
the desired match is difficult to ensure at higher frequencies 
where small differences in the mechanical structure and
feed-point connections can cause noticeable mismatch. Such mismatch,
if any, would manifest in distortion of the resultant beam-shape.
Except in one of the bands (i.e. at 630 MHz, and in one of the 
polarizations, as seen in second row's first sub-panel of
Figure~\ref{fig_monitor_time}), such 
distortion is not apparent, indicating reasonable match within
the array pair in all other bands, including those at higher
frequencies.
\par
A closer look at the beam width estimates reveals
that they deviate significantly from the mean trend at the two shortest
wavelengths (Figure~\ref{fig_beamwidths}), possibly 
implying the illuminated area to be effectively smaller by a 
factor of about 2 at these wavelengths.
Another potential cause for the loss in sensitivity is related 
to signal-to-noise ratio (S/N) in different bands viewed at the stage
of optical fiber transmission. As already mentioned, 
spectral dependence of the amplifier gains and cable losses 
together causes significant gradient in the signal level across
our wide-band, although we have tried to reduce it as much as possible. 
Due to constraints on the total RF power level 
to be input to the fiber transmission unit (after allowing for 
due headroom for RFI), the contrast in spectral density of our 
signal to that of the transmitter noise gets seriously
compromised in bands where signal level is relatively low 
(resulting from gain-slopes). Improvement in sensitivity at the
higher bands would necessarily require addressing both these issues,
with suitable modifications in the feed design as well as employing
multiple fibers to carry different sections (2 to 3) of the wide band
separately to avoid degradation of S/N. While these are being pursued,
efforts are also going on to miniaturize the whole receiver, 
significantly reducing its physical size for ease of portability.
\par
For completeness, we mention infrequent but noticeable occurrences
of data-slips, caused by occasional slowness in capturing of the 
high-rate stream of packets to the disk, in the present data
(such gaps in data sequence are apparent in Figure~\ref{fig_dyn_spec};
note the randomly distributed dark vertical lines across various bands).
This issue has now been resolved 
by using a software tool ``Gulp'' \citep{Corey07} which provides a 
significantly larger buffer between packet-capturing from the 
ethernet and disk-writing, enabling lossless capture of packets at high rates.
\par
Although the development of the MBR is primarily motivated by the 
tomographic studies of pulsar polar emission regions and would aid
probe of several aspects of pulsar emission at single-pulse resolution, 
the potential of simultaneous multi-frequency observations facilitated 
by such a system can not be overstated. In particular, while looking for
fast transients, simultaneous view in multiple bands is 
of tremendous advantage in decisively discriminating the signals 
of astronomical origin from those due to RFI. While the
enhancement in the formal sensitivity due to the increase in
available bandwidth can be quantified trivially, 
the added advantage of being able to verify the consistency of a 
dispersed signature of any candidate event across such a wide band,
though crucial, is not easily quantifiable. 
Apart from this, in many other multi-frequency single-dish studies 
where simultaneity of observations may not be crucial,
usage of the MBR can help save telescope time, and the implicit compatibility 
in data spans, types and formats, etc, in multiple frequency bands would 
offer a desired ease in off-line data processing and comparisons. 
This capability would benefit in studying spectral evolution of the 
{\it average} properties of pulsars (e.g. pulse intensity and shapes in 
full Stokes), and propagation effects (including interstellar 
scattering/scintillation), as well as in several single-dish continuum 
studies (including polarization), in general. 
The MBR also offers a significant tunability in the center frequencies
of each of the 16-MHz wide sub-bands separately, a feature 
advantageous particularly for surveys/studies of recombination lines. 
\par
In summary, the design and development of a 
self-contained multi-band receiver (MBR), reported in this paper,
opens up a new powerful mode of 
simultaneous multi-frequency observations in 10 bands when
used with a sufficiently large aperture. The design details of
several components (particularly that of the multi-band feed)
could be of relevance to on-going and future developments in radio
astronomy instrumentation,
for example, those related to the SKA and FAST.
The design and structure of the MBR is generic enough to enable 
its use at any suitably large single-aperture telescope. 
The preliminary results, from the initial observations carried out 
using our MBR system at the GBT, provide a glimpse of the exciting
phase ahead in such pulsar emission tomographic studies we intend
to carry out in greater detail.
\section*{Acknowledgments}
We are grateful to our colleagues at 
RRI 
(\textit{Peeyush, C. R. Subrahmanya, Girish, Kamini, Raghunathan,
Chandrashekara, Prabu, Rajgopal, Wences Laus, Mamatha T. S., Vishakha, 
Sohan, Ramanna, Muniraju,
Ateequlla, Venu, Achankunju, Srinivas, Dhamodran, Suresh, Gokul,
Mohandas, Gopal, Sunand, Elumalai, Sunderaj, Paul, Puttaswamy, Sivasakthi,
Abdul Majeed, Ananda, Jacob, Sridhar, Nandakumar, Srinivasa Murthy, 
Ramamurthy, P. V. Subramanya, Mamatha Bai, Krishnamaraju, Marisa, Vidya, 
Shashi and other colleagues}), 
IIA
(\textit{Kathiravan, Ramesh, Rajalingam and other team members at 
Gauribidanur}), 
GMRT
(\textit{Rajaram, Praveen, Yashwant, Kale, Nagarathnam, 
Suresh Sabhapathy, Ishwar Chandra, Nimisha,
the telescope operators and the GMRT operations team}), 
GBT 
(\textit{Steve White, Rusty Taylor, Bob Anderson, Harry Morton, Bob Simon, 
Chris Clark, Pete Chestnut, Frank Ghigo, Rick Fisher, Chuck Niday, 
Wolfgang Baudler, Jessica Thompson, Sherry Sizemore, Christine Plumley,
Eric Knapp, Kevin Gum, Dave Rose, Barry Sharp, Donna Stricklin, Greg Monk, 
the telescope operators and the Operations \& Mechanical teams})
and Angela \& Jeff Hegan for their generous support and help at various 
stages of this work. NM, DBD, HS and STV gratefully acknowledge financial
support from RRI under the visiting student programme.
GMRT is an international facility run by the National Centre for Radio 
Astrophysics of the Tata Institute of Fundamental Research.
The National Radio Astronomy Observatory is a facility of the National 
Science Foundation operated under cooperative agreement by Associated 
Universities, Inc.\\
{\it Facilities:} \facility{GMRT}, \facility{GBT}

\begin{table}
\caption{RF Filter Specifications}
\begin{tabular}[c]{lcc}
\hline
\hline
S.No.&Center Frequency&Band-width\\
&(MHz)&(MHz)\\
\hline
1	& 119 & 20 \\ 
2	& 172 & 21 \\ 
3	& 232 & 37 \\ 
4	& 330 & 38 \\ 
5	& 425 & 46 \\ 
6	& 629 & 59 \\ 
7	& 730 & 62 \\ 
8	& 825 & 82 \\ 
9	& 1200 & 112 \\ 
10	& 1450 & 104 \\
\hline
\end{tabular}
\label{table_filters}
\end{table}
\label{lastpage}
\end{document}